# $As_2S_3$, $As_2Se_3$ and $As_2Te_3$ nanosheets: Superstretchable semiconductors with anisotropic carrier mobilities and optical properties

Bohayra Mortazavi*[a,b], Fazel Shojaei[c], Maryam Azizi[c], Timon Rabczuk[d] and Xiaoying Zhuang**[d]

[a]*Institute of Continuum Mechanics, Leibniz Universität Hannover, Appelstraße 11, 30157 Hannover, Germany.*
[b]*Cluster of Excellence PhoenixD (Photonics, Optics, and Engineering–Innovation Across Disciplines), Gottfried Wilhelm Leibniz Universität Hannover, Hannover, Germany*
[c]*Schoolof Nano Science, Institute for Research in Fundamental Sciences (IPM), 19395-5531 Tehran, Iran.*
[d]*College of Civil Engineering, Department of Geotechnical Engineering, Tongji University, Shanghai, China.*

**Abstract**

In this work, density functional theory calculations were carried out to explore the mechanical response, dynamical/thermal stability, electronic/optical properties and photocatalytic features of monoclinic $As_2X_3$ (*X*=S, Se and Te) nanosheets. Acquired phonon dispersions and ab-initio molecular dynamics results confirm the stability of studied nanomembranes. Observation of relatively weak interlayer interactions suggests that the exfoliation techniques can be potentially employed to fabricate nanomembranes from their bulk counterparts. The studied nanosheets were found to show highly anisotropic mechanical properties. Notably, new $As_2Te_3$ 2D lattice predicted by this study is found to exhibit unique superstretchability, which outperforms other 2D materials. In addition, our results on the basis of HSE06 functional reveal the indirect semiconducting electronic nature for the monolayer to few-layer and bulk structures of $As_2X_3$, in which a moderate decreasing trend in the band-gap by increasing the thickness can be established. The studied nanomaterials were found to show remarkably high and anisotropic carrier mobilities. Moreover, optical results show that these nanosheets can absorb the visible light. In particular, the valence and conduction band edge positions, high carrier mobilities and optical responses of $As_2Se_3$ nanosheets were found to be highly desirable for the solar water splitting. The comprehensive vision provided by this study not only confirm the stability and highly attractive electronic and optical characteristics of $As_2S_3$, $As_2Se_3$ and $As_2Te_3$ nanosheets, but also offer new possibilities to design superstretchable nanodevices.

Corresponding authors: *bohayra.mortazavi@gmail.com; **zhuang@ikm.uni-hannover.de

# 1. Introduction

Among various classes of materials, two-dimensional (2D) nanomaterials can be currently considered as the most vibrant family, in which new members are continuously being theoretically predicted and experimentally realized. Outstanding attentions to this group of materials originates from the wide range of application prospects and also the unique properties of its members, with exceptional abilities to exhibit diverse and contrasting properties. Graphene [1–3], the 2D form of carbon atoms, also called as the wonder material is the most prominent member of 2D materials, which shows a unique ability to simultaneously exhibit remarkable mechanical stiffness [4], ultrahigh thermal conductivity [5] and exceptional optical and electronic features [3,6–8]. Graphene is already stepping into some critical technologies, such as; nanoelectronics, optoelectronics and aerospace industry. Graphene can be also considered as the representative member of 2D materials with highly symmetrical and isotropic lattices, which includes also other well-known compositions, like; hexagonal boron-nitride (h-BN) [9,10], silicene [11,12], germanene [13], indium selenide [14] and 2H and 1T phases of transition metal dichalcogenides [15,16]. The 2D materials diversity can be also seen in the broad range of atomic structures, with large number of compositions showing anisotropic lattices, such as; borophene [17,18], phosphorene [19–21] and few very low-symmetry systems, like 1T' phases of transition metal dichalcogenides [22,23] and germanium phosphide [24]. By having a glance on the existing rich literature, it is conspicuous that the majority of researches have been devoted to fabricate and explore the properties of isotropic and symmetrical 2D lattices. During the past few years, however anisotropic 2D lattices like; borophene and phosphorene have been among the most attractive and active fields of researches in 2D materials. In such a trend, most recently low-symmetry 2D systems are also gaining remarkable attention, in order to design novel devices, in which the conventional and highly symmetrical materials fail to deliver the desirable performance. This new trend is due to the fact that low-symmetry materials can yield highly anisotropic properties and these features can be accordingly utilized in order to engineer direction-dependent systems, such as polarized optics and photodetectors, digital inverters, and artificial synaptic devices [24–27].

In recent years, arsenic based layered materials have gained remarkable attentions for the applications in electronic devices, sensors, and energy systems [28–30]. $As_2S_3$ and $As_2Se_3$ with monoclinic structures are bulk-layered materials, which belong to the anisotropic material family. Recently, $As_2S_3$ in the 2D form was experimentally realized by Siskins *et al.* [31], which shows outstanding chemical stability and highly anisotropic mechanical and optical properties. Motivated by this latest experimental advance, in this work our objective is to explore the dynamical and thermal stability, mechanical response, electronic and optical characteristics of monoclinic $As_2X_3$ ($X$=S, Se and Te) nanosheets. In particular, we elaborately analyze the direction dependency of the mechanical and optical properties of these novel nanosheets. Worthy to mention that among all areas for the application of 2D materials, energy storage and conversion systems are among the most appealing and attractive ones. In particular, the production of hydrogen fuel via the solar water splitting and design of advanced rechargeable metal-ion batteries, using the 2D materials as active building blocks have been extensively explored during the last decade [32]. The tremendous interests toward the employment of 2D materials in the energy storage/conversion systems originate from their large surface to volume ratio, outstanding mechanical flexibility, remarkably high electron mobility and chemical and thermal stabilities. Therefore, in this study we particularly evaluate the suitability of $As_2X_3$ ($X$=S, Se and Te) nanomembranes for the overall solar water splitting, with respect to their electronic band structure, carrier mobilities and optical characteristics.

## 2. Computational methods

Geometry relaxations and electronic/optical structure calculations in this work were all performed via density functional theory (DFT) calculations using the *Vienna Ab-initio Simulation Package* (VASP) [33–35]. We employed the generalized gradient approximation (GGA) and Perdew−Burke−Ernzerhof (PBE) [36] method with a plane-wave cutoff energy of 500 eV. We also adopted the dispersion scheme of DFT-D2 [37] to more accurately take into account the Van der Waals (vdW) interactions. The convergence criterion for the electronic self-consistent-loop was set to be $10^{-4}$ eV. For the ab-initio calculations periodic boundary conditions were applied along all three Cartesian directions. For the monolayer systems the simulation box size along the sheets normal direction (z direction) was set to 20 Å to avoid the interactions with adjacent layers. For the multi-layered structures this length was accordingly increased so that no adjacent layer artifacts occur. The geometry optimizations were achieved using the conjugate gradient method with the a convergence criterion of 0.01 eV/Å for Hellmann-Feynman forces, in which a 7×7×1 Monkhorst-Pack [38] k-point mesh size was used. Uniaxial tensile simulations were carried out to evaluate the mechanical properties, by applying the strain levels gradually. In this case in order to ensure the satisfaction of unidirectional stress conditions, after applying the strain levels and subsequent conjugate gradient geometry optimization, the box size along the transverse direction of loading was adjusted to reach negligible stress values (less than 0.02 N/m). Since PBE systematically tends to underestimate the band gaps of semiconductors and insulators due to the self-interaction errors, screened hybrid functional of HSE06 [39] was employed to more accurately estimate the electronic structures.

In order to calculate the second order force constants, density functional perturbation theory (DFPT) simulations were performed over 2×5×1 super-cells. We then employed PHONOPY code [40] in which the DFPT results were used as the input in order to acquire phonon dispersions. Ab-initio molecular dynamics (AIMD) simulations were conducted with a time step of 1 fs, over 2×3×1 super-cell structures, using a 2×2×1 k-point mesh size. Optical calculations were performed on the basis of Bethe Salpeter and Casida equation [41] constructed over the HSE06 results, as implemented in the VASP using 3×7×1 k-point grid. Carrier mobility for the considered 2D systems were calculated from the deformation potential approximation [42] via: $e\hbar^3 C_{2D}/KTm_e^* m_d (E_L^i)^2$, where $K$ is the boltzmann constant, $\hbar$ is the Planck constant, $C_{2D}$ is the elastic modulus along the transport direction, $m^*$ is the effective mass of the carrier in the same direction, $m_d$ is the average effective mass in the two directions and $E_L^i$ mimics the deformation energy constant of the carrier due to phonons for the i-th edge band along the transport direction.

### 3. Results and discussions

Energy minimized $As_2S_3$, $As_2Se_3$ and $As_2Te_3$ monolayers with rectangular primitive unit-cells, monoclinic lattices and Pmn21 space group are illustrated in Fig. 1. Each primitive unit-cell contains two formula units of $As_2X_3$ (*X*=S, Se and Te) obeying the octet rule, in which As atoms are three coordinated (three As-*X* bonds) and X atoms are two coordinated (two As-*X*). This bonding pattern results in nanoporous sheets that are made exclusively of As-X bonds. It is shown, these structures show anisotropic atomic lattices. The lattice constants of bulk $As_2S_3$ along the *x* and *y* directions, were experimentally measured to be 11.46 and 4.22 Å, respectively [43]. These values are very close to our predicted corresponding values of 11.359 and 4.456 Å, respectively, for the single-layer $As_2S_3$, on the basis of PBE and DFT-D2 [37] vdW dispersion correction. Worthy to note that with the pure PBE the corresponding lattice constants for the $As_2S_3$ monolayer were found to be 11.391 and 4.607 Å, respectively. While the predicted lattice constants along the elongated direction (*x* direction) are very close to that of the bulk structure with and without vdW dispersion correction, it is apparent that the incorporation of dispersion correction yields more accurate estimation for the lattice length along the other in-plane direction (*y* direction). By increasing the weight of chalcogen atoms in the considered monolayers, the lattice constant along the elongated direction (*x* direction in this work) increases, whereas the lattice length along the perpendicular direction (*y* direction in this work) remains almost unaffected. To facilitate the future studies, the energy minimized single- and double-layer lattices in the VASP native format are included in the supporting information document. In Fig, 1 the electron localization function (ELF) [44] within the unit-cells are plotted to investigate the bonding mechanism in these nanosheets. We remind that ELF is a three-dimensional function with a value ranging between 0 and 1. Generally ELF values close to 1 indicate strong covalent bonding or lone pair electrons, while smaller ELF values ($\leq 0.5$) correspond to metallic or ionic bonds, and low electron density localization. It can be clearly seen that electron localization is happening around the center of arsenic and chalcogen atoms, which can be assigned to their lone pair electrons. Rather smaller, but still large electron localization is observed in between arsenic and chalcogen atoms along the As-*X* bonds, implying the dominance of covalent interactions in the considered nanosheets. As it is clear from the side views shown in Fig. 1, by increasing the weight of chalcogen atoms the ELF contour around the center of chalcogen and arsenic bonds enhance. This can be an indication for the improvement of the covalent bonding in these systems.

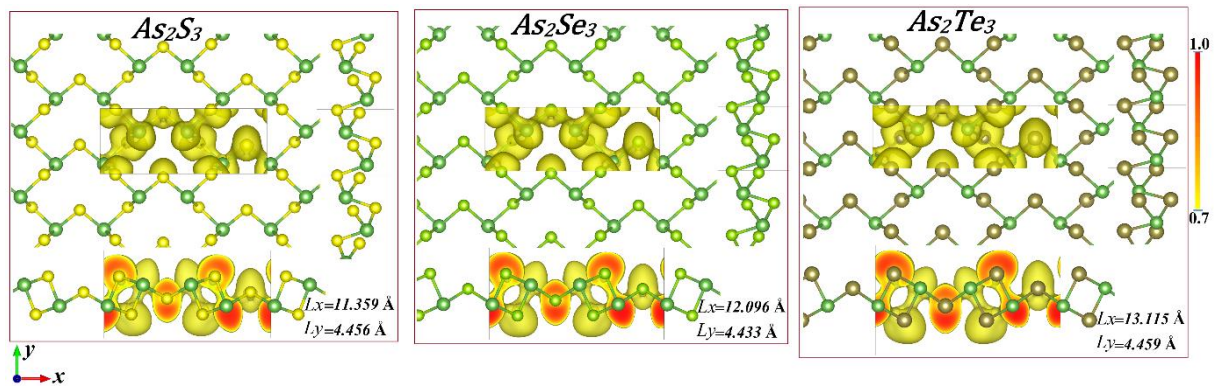

**Fig. 1**, Top and side views for the atomic structure of $As_2S_3$, $As_2Se_3$ and $As_2Te_3$ monolayers. Contours illustrate the electron localization function [44] within the unit-cell. The iso-surface level for ELF is set as 0.7.

As a common approach to examine the stability of novel 2D materials, we first study the dynamical stability of $As_2S_3$, $As_2Se_3$ and $As_2Te_3$ monolayers by plotting the phonon dispersion relations, as shown in Fig. 2. As the characteristic feature of the 2D materials, the phonon dispersions present three acoustic modes, two of them with linear dispersion, and the remaining one with a quadratic relation. The acquired phonon dispersion for the $As_2Te_3$ monolayer reveals the absence of negative branches which clearly confirm the dynamical and structural stability. On the other side for the As$_2$S$_3$ and As$_2$Se$_3$ monolayers, in one of the acoustic modes around the gamma point, negligible pit shaped negative branches are observable with the maximum frequency smaller than 0.1 THz. The negligible U shaped feature in the beginning of the gamma-Y and gamma-S paths are the signature of the flexural acoustic mode, which are usually hard to converge in the 2D materials [45] and can be removed by increasing the accuracy of calculations, upon the increasing the super-cell size or k-point grid. On the other hand, in the cases of multi-layer structures or supporting over a substrate, these slight U shaped negative branches normally vanish. As expected by increasing the weight of chalcogen atoms in the considered systems, the frequency ranges suppress, which can normally lead to lower thermal conductivity. According to the acquired phonon dispersion results shown in Fig. 2, it can be concluded that the As$_2$S$_3$, As$_2$Se$_3$ and As$_2$Te$_3$ nanosheets are dynamically stable. For the assessment of practical applications in devices, thermal stability is another critical factor that should be carefully examined. According to AIMD results illustrated in Fig. S1 and Fig. S2, it can be seen that after 13 ps of annealing at the elevated temperature of 500 K, all chemical bonds in these systems were kept completely intact, and thus confirming the thermal stability of As$_2$S$_3$, As$_2$Se$_3$ and As$_2$Te$_3$ nanosheets. Phonon dispersions and AIMD results reveal the desirable thermal and dynamical stability of considered nanosheets for the practical application.

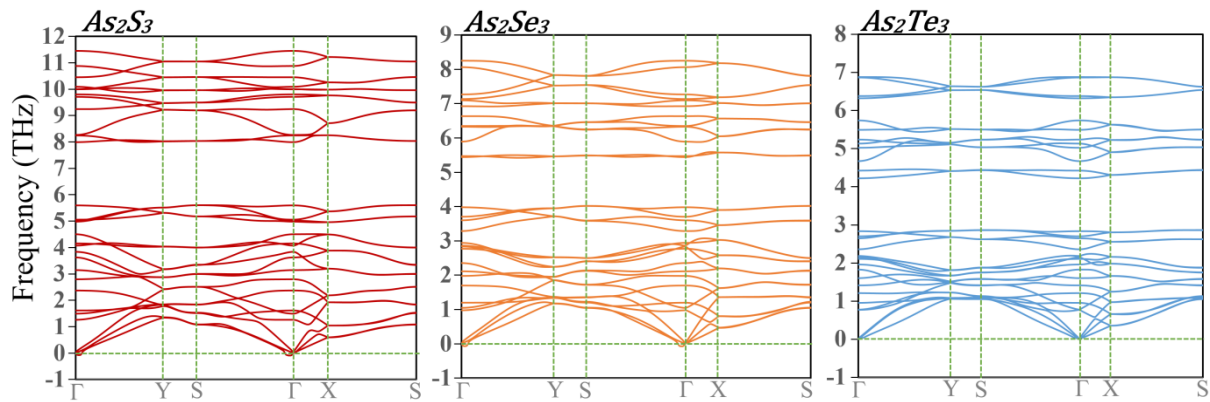

**Fig. 2**, Phonon dispersion results for the single-layer As$_2$S$_3$, As$_2$Se$_3$ and As$_2$Te$_3$.

We next investigate the feasibility of isolation of single- or few-layer As$_2$X$_3$ (*X*=S, Se and Te) nanomembranes from their multi-layered or bulk counterparts. In order to address this important property, we evaluate the cleavage energies for the exfoliation process and compare the predicted energies with other well-known successfully isolated monolayers. As a common approach in this case we considered six-layered slab for every structure, in which the top layer was gradually moved toward the vacuum and the change in the energy was recorded. Worthy to note that for the construction of six-layered slabs we used the same stacking pattern as that in the bulk arsenic trichacogenides systems. As shown in Fig. 3 results, for the all considered systems after the separation distance of 6 Å the cleavage energy convincingly converges and for higher separation distances the effects of Van der Waals interactions become negligible. According to our simulations the cleavage energies for $As_2S_3$, $As_2Se_3$ and $As_2Te_3$ monolayers were calculated to be 0.23, 0.33 and 0.52 J/m$^2$, respectively. Notably, the cleavage energies for $As_2S_3$ and $As_2Se_3$ nanosheets are smaller than the experimentally calculated cleavage energy for graphene (0.37 J/m2) [46], which highlight the feasibility of fabricating these monolayers from their bulk counterparts. On the other side, to the best of our knowledge to date the $As_2Te_3$ multi-layered systems with the monoclinic lattice have not been reported. Therefore, for this structure the chemical growth techniques should be employed to fabricate nanomembranes. Nonetheless, although the exfoliation energy for $As_2Te_3$ nanosheets is by around 40% higher than that of the graphene, it is yet almost half of that for experimentally realized $Ca_2N$ monolayer (1.09 J/m$^2$) [47].

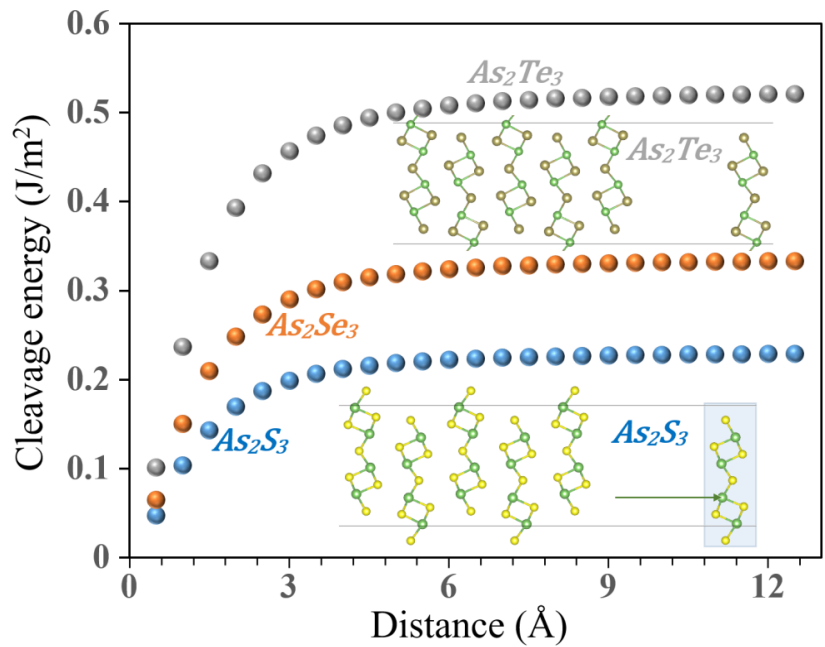

**Fig. 3**, Cleavage energy as a function of separation distance between the $As_2X_3$ monolayers and corresponding five-layered slabs showing convergences after the distance of 6 Å.

Next, we explore the mechanical properties of $As_2S_3$, $As_2Se_3$ and $As_2Te_3$ nanomembranes according to the uniaxial tensile results. Because of the anisotropic nature of studied nanomaterials, uniaxial tensile simulations were conducted along the $x$ and $y$ planar directions. Because of the layered structures and the existence of vacuum along the sheets normal direction, after the relaxation process the stress along the normal direction of the sheet is negligible. Nonetheless, after the application of loading strain normally the stress in the perpendicular direction of loading evolve due to the Poisson's effect. This way to satisfy the unidirectional stress condition, the box size along the perpendicular direction of loading was altered such that after the relaxation the aforementioned effect gets compensated and the absolute value of stress stay within a negligible range (less than 0.03 N/m). In Fig. 4 the estimated uniaxial stress-strain responses of $As_2S_3$, $As_2Se_3$ and $As_2Te_3$ nanosheets along $x$ and $y$ directions are compared. As it is clear, the stress-strain relations are completely different along the two different loading directions, due to the highly anisotropic nature of the studied nanosheets. These stress-strain curves follow the usual trends observable in other conventional materials, as they start with an initial linear relation and after reaching a yield point they continue by a nonlinear response. We remind that the slope of initial linear section in the stress-strain can be used to extract the elastic modulus. It is interesting that the initial parts of the stress-strain curves for the considered nanosheets are very close, which reveals that the type of chalcogen atoms in these lattices does not considerably affect the elastic modulus. The elastic modulus of $As_2S_3$, $As_2Se_3$ and $As_2Te_3$ along the $x$ direction was measured to be 29.3, 28.7 and 27.3 N/m, respectively, which are by around a factor of three higher than those along the $y$ direction, 8.3, 9.0 and 9.9 N/m, respectively. On the other side, it is clear that for the uniaxial loading along the $x$ direction, the initial linear response covers the majority of stress-strain curve, whereas for the loading along the $y$ direction the nonlinear section becomes more pronounced. These initial observations reveal that along the $x$ direction these nanomaterials are elastic and brittle, whereas along the $y$ direction they show superstretchability, likely to rubber. The predicted stress-strain relations also reveal that along the $x$ direction these nanosheets show distinctly higher tensile strengths as compared with the $y$ direction. The tensile strength of $As_2S_3$, $As_2Se_3$ and $As_2Te_3$ along the $x$ direction was found to be 5.7, 4.6 and 3.4 N/m, respectively, which are by more than two folds higher than those along the $y$ direction, 2.3, 2.0 and 1.7 N/m, respectively. Unlike the elastic modulus, it is clear that by increasing the weight of the chalcogen atoms in the $As_2X_3$ monolayers the tensile strength decreases monolithically.

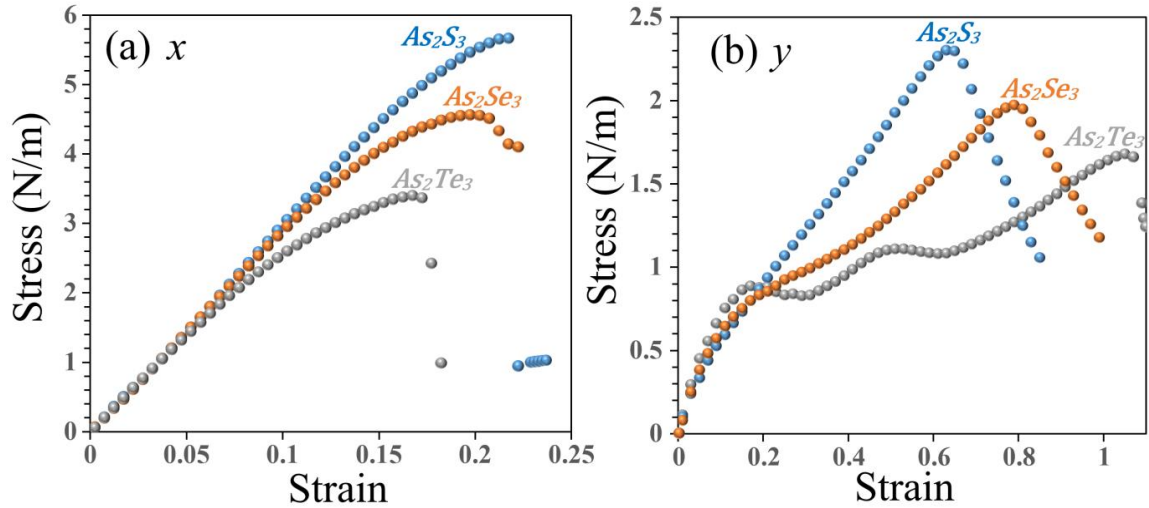

**Fig. 4**, Mechanical responses of As$_2$X$_3$ monolayers for the uniaxial loading along the (a) *x* and (b) *y* directions.

The nature of highly anisotropic mechanical properties of considered nanosheets can be also well explained by the predicted stress-strain relations. Basically the bond elongations during the loading condition result in the sharp increase in the stress values. However, the structural deflections such as the bond rotations and the contraction of the sheets along the perpendicular direction of loading can facilitate the deformation and result in smoother increase in the stress values. Highly linear and brittle behavior for the uniaxial loading along the *x* direction suggests that for this loading the deformation is dominated by the bond elongation. Since the stretchability of a chemical bond is an intrinsic property and thus limited, the structures fail at relatively low strain levels. On the other side, for the loading along the *y* direction the elastic modulus is much lower than the *x* direction, which implies that from the early stages of the loading the deformation is not only achieved by the bond elongation but also with structural deflection. After the initial linear response and especially for the cases of As$_2$Se$_3$ and As$_2$Te$_3$ nanosheets, the stress values evolve very smoothly which suggest that the structural deflections outperform the bond elongations during the deformation process.

This anisotropic deformation mechanism in the considered nanosheets can be also explained by considering the atomic structures at different strain levels, as compared in Fig. 5. It is clear that even for the loading along the *x* direction, during the loading the sheet contracts considerably along the width. Such that the structural deflections also happen for the loading along this direction, though the bond elongation yet dominates the deformation. In this case, the As-*X*-As bonds that are oriented along the loading undergo remarkable stretching and subsequently the final rupture occurs by the breakage of As-*X* bonds. This way, the lower tensile strength by increasing the weight of chalcogen atoms confirms that the strength of As-$X$ bonds decreases accordingly. When comparing with the loading along the $y$ direction, due to the unidirectionality of the lattice porosity with the stretching direction, there exists more available space for the sheet contraction during the deformation. Beside this, some As-$X$ bonds during the deformation and sheets contraction become oriented along the loading direction. In this case the chalcogen atoms can behave like pivots which remarkably facilitate the deformation and stretchability as well. To provide a better comparison, we remind that the stretchability of pristine graphene and hexagonal boron-nitride are around ~0.27 [48] and 0.3 [49], respectively. This way, the stretchability of $As_2S_3$ and $As_2Te_3$ are by more than two and three folds larger than graphene, respectively. To the best of our knowledge, Boron-graphdiyne with the graphene-like and highly porous atomic lattice yields the highest stretchability of 0.88, which outperforms other 2D materials [50]. It is amazing that the much denser $As_2Te_3$ nanosheet outperforms Boron-graphdiyne with respect to the stretchability. This highly attractive observation can be attributed not only to the nanoporosity in the atomic lattice and the possibility of the $As_2X_3$ structures to contract considerably during the loading, but also to the pivot like behavior of chalcogen atoms in these systems. As the interests toward the stretchable and wearable electronics are continuously increasing, our findings may serve as useful guides to guide the design of other superstretchable 2D semiconductors.

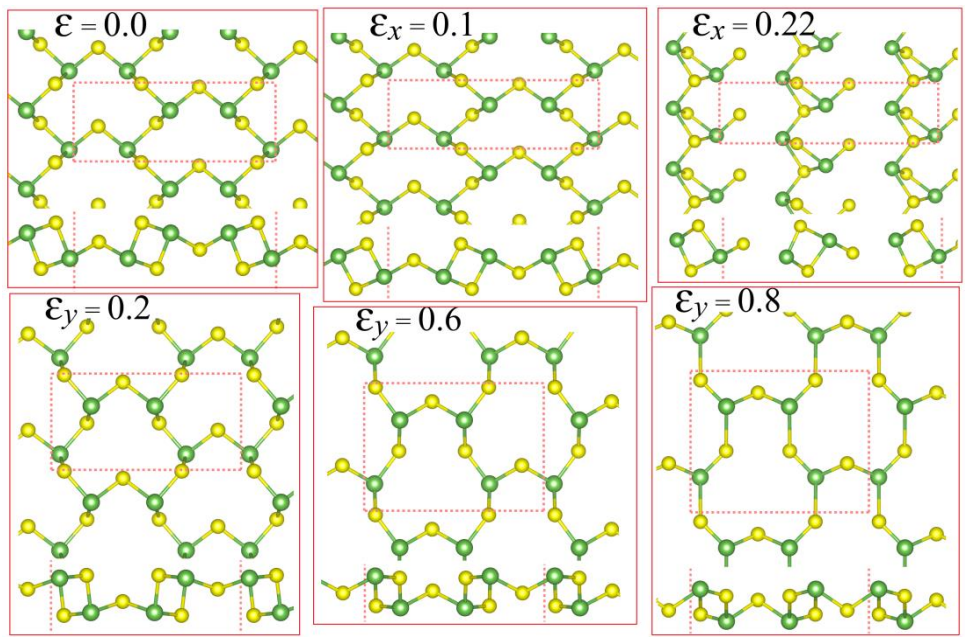

**Fig. 5**, Top and side views of the $As_2X_3$ monolayers at different strain levels. $\varepsilon_x$ and $\varepsilon_y$ depict the strain levels for the uniaxial loading along the $x$ than $y$ directions, respectively.

We next examine the electronic properties of arsenic trichalcogenide nanosheets by calculating their HSE06 electronic band structures. Fig. 6 shows the electronic band structures and partial density of states (PDOS) of the monolayer and bilayer $As_2X_3$. We first focus on the electronic properties of the single-layer structures. One can clearly see that $As_2X_3$ monolayer are semiconducting with indirect band gaps of 3.277 eV for $As_2S_3$, 2.596 eV for $As_2Se_3$, and 1.774 eV for $As_2Te_3$ monolayer on the basis of HSE06 functional. Worthy to note that in an earlier study by Debbichi and co-workers [43], the band gap of $As_2S_3$ and $As_2Se_3$ monolayers were reported to be 2.95 and 2.25 eV, respectively, which are by around 10% smaller than our estimations. The discrepancy between our results can be attributed to the fact that in the aforementioned study authors did not consider the vdW dispersion correction, which as discussed earlier result in the overestimation of lattice constant along the *y* direction. According to the both HSE06 and PBE (find Fig. S3) functional for the studied monolayers, the valance band maximum (VBM) occurs at certain *k*-points on *Γ-Y* path, while the conduction band minimum is located at *Γ* point. Worthy to also mention that our results for the electronic band structures within the PBE functional (shown in Fig. S3) suggest that the spin-orbit coupling (SOC) is convincingly negligible in the studied nanosheets. It is conspicuous that the band gap decreases with increasing the weight of chalcogen atom.

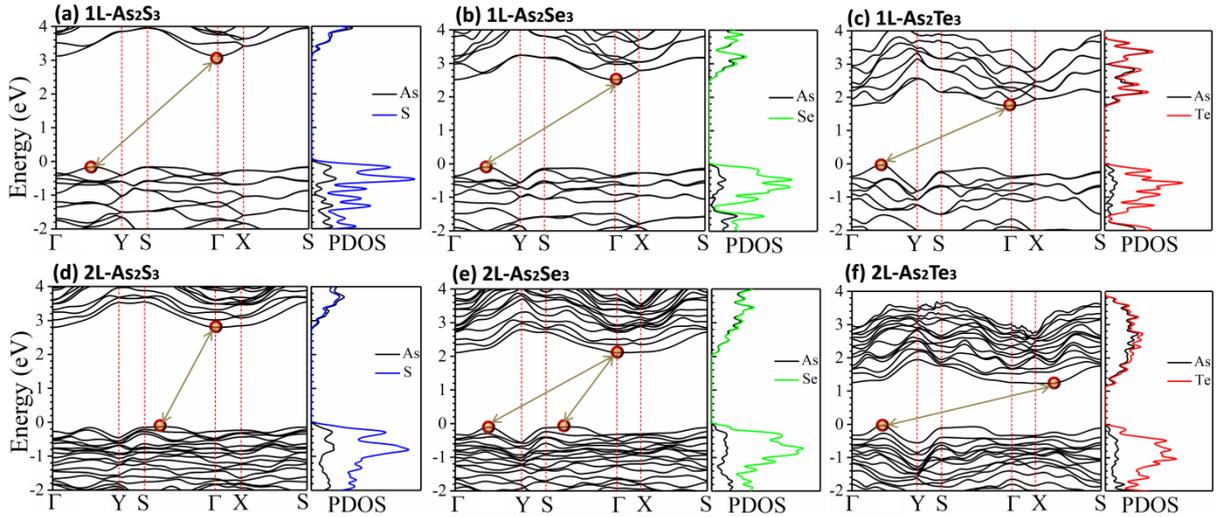

**Fig. 6**, Electronic band structures and total electronic density of states (DOS) of single-layer (1L) and bilayer (2L) $As_2S_3$, $As_2Se_3$ and $As_2Te_3$ predicted by the HSE06 functional. The highlighted points show the locations of CBM and VBMs.

We calculated the charge density distributions at VBM and CBM of the studied monolayers and the acquired results for $As_2S_3$ monolayer are shown in Fig. 7 (find Fig. S5 for $As_2Se_3$, and $As_2Te_3$ lattices). Our results for PDOS and charge density distributions indicate that VBM is mainly derived from lone pair electrons of chalcogen atoms with a minor contribution of antibonding (As-$X$)$_{in\text{-}plane}$ σ* states. It is apparent that lone pair electrons of larger atoms in a group are found at higher energy levels. Consequently, the VBM moves upward when $X$ changes from S to Te and consequently resulting in smaller band-gaps. On the other hand, the CBM is equally contributed by $p$ orbital of As and $X$ atoms, representing bonding (As-$X$)$_{in\text{-}plane}$ π and antibonding (As-$X$)$_{vertical}$ σ* states.

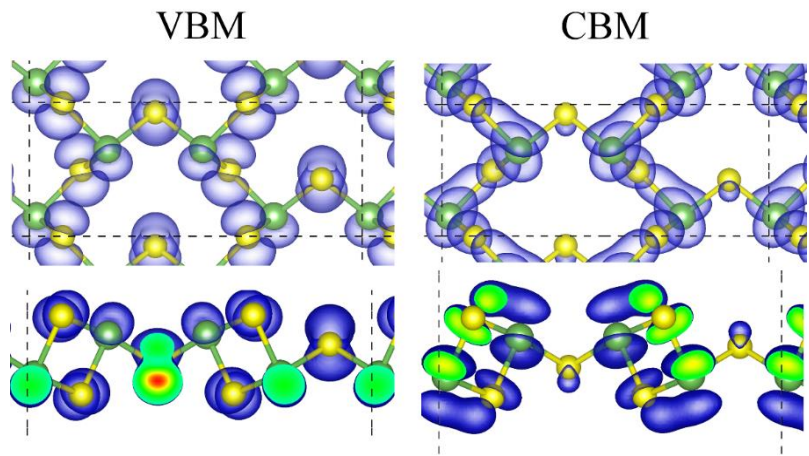

**Fig. 7**, Top and side views for the calculated charge density distributions of VBM and CBM states of As$_2$S$_3$ monolayer using the HSE06 method. Due to the similarity in the electronic structure of As$_2$S$_3$, As$_2$Se$_3$, and As$_2$Te$_3$ monolayers, the results for only As$_2$S$_3$ monolayer are shown here and those for other two monolayers are presented in the Fig. S5. The iso-surface value is set to 0.003 e/Å$^3$.

The electronic properties of these 2D materials also exhibit sensitivity with respect to the number of layers, due to the quantum size confinement effects. To examine the quantum confinement effects in arsenic trichalcogenides, we also calculated the electronic band structures of their bilayer and bulk counterparts. We used the same stacking pattern as in bulk arsenic trichacogenides to construct the bilayer systems. It is found that although in a few cases the transition k-points changed compared to those of the monolayer counterparts, the indirect band-gap nature was found to be intact for bilayer and bulk systems (shown in Fig. S3). For the bilayer As$_2$S$_3$, As$_2$Se$_3$ and As$_2$Te$_3$ the band-gaps were found to be 2.928, 2.202 and 1.252 eV, respectively, which are close to those of the bulk lattices, 2.865, 1.958 and 0.937 eV, respectively. As it is clear, the band-gap drops somewhat considerably from the monolayers to bilayers, but from the bilayer systems the effects of quantum confinement suppress. Our analysis of charge density distributions of bilayer systems indicates that their VBMs are made of antibonding interaction between the VBMs of the two layers, while the CBMs come from bonding interaction between the CBMs of the two layers. Consequently, the VBM energetically shifts upward and CBM moves downward resulting in smaller band gaps than their monolayer counterparts. In the future studies, the effects of mechanical straining, defects and impurities [51–58] can be studied on engineering of the electronic properties of these novel 2D systems.

It is well-known that hydrogen fuel can yield exceptionally high energy density among all types of fuels, which highlights the promising application prospect of producing this fuel via the solar water splitting methods. Because of the semiconducting electronic nature of studied nanosheets, we next examine their application prospect for the overall solar water splitting, which involves two redox half reactions:

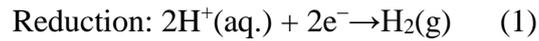
$$\text{Reduction: } 2H^+(aq.) + 2e^- \rightarrow H_2(g) \quad (1)$$

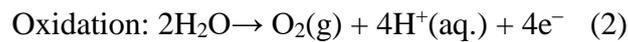
$$\text{Oxidation: } 2H_2O \rightarrow O_2(g) + 4H^+(aq.) + 4e^- \quad (2)$$

At pH equal to zero, the $H^+/H_2$ reduction potential and the $O_2/H_2O$ oxidation potential are −4.44 and −5.67 eV, respectively [59,60]. On this basis the minimum band-gap for this process is 1.23 eV, which can be satisfied by all three studied nanosheets in this work. Nevertheless, for the successful water splitting the exact positions of CBM and VBM should be also taken into account. In another word, a semiconductor can be suitable for the water splitting only if it yields a VBM less negative than the $H^+/H_2$ reduction potential and a CBM edge more negative than the $O_2/H_2O$ oxidation potential. To explore the satisfaction of aforementioned aspects, the CBM and VBM positions for the monolayer and bilayer $As_2S_3$, $As_2Se_3$ and $As_2Te_3$ nanpsheets with respect to the vacuum level (set at 0 eV) along with the $H^+/H_2$ reduction and $O_2/H_2O$ oxidation energy levels for the water splitting at pH=0 and 7 are compared in Fig. 8. As it can be seen $As_2S_3$ and $As_2Se_3$ nanosheets completely satisfy the mandatory band positions, while $As_2Te_3$ counterparts fail for the $H^+/H_2$ reduction process, though they are yet suitable for $O_2/H_2O$ oxidation. By considering the energy levels for the reduction and oxidation processes, it can be seen that the $As_2Se_3$ nanosheets yield closer values to the energy limits and thus they are more suitable for the water splitting than $As_2S_3$ counterparts. Our results nonetheless suggest that $As_2S_3$ and $As_2Se_3$ nanosheets offer suitable electronic structure for the overall solar water splitting under various environments with different pHs.

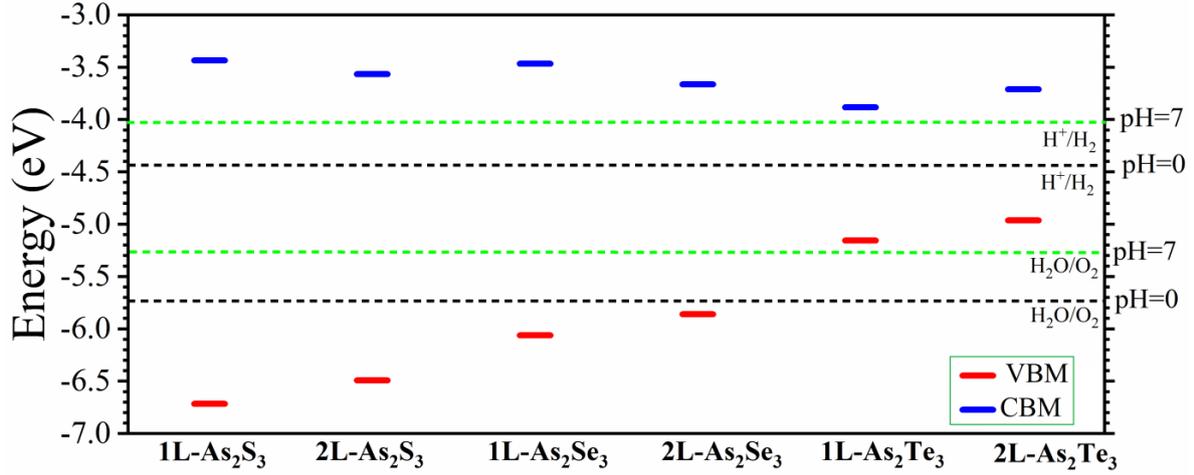

**Fig. 8**, CBM and VBM positions with respect to the vacuum level (set at 0 eV) for the monolayer (1L) and bilayer (2L) $As_2S_3$, $As_2Se_3$ and $As_2Te_3$ nanosheets estimated by HSE06 method, along with the reduction and oxidation energy levels for the water splitting at pH equal to zero and seven.

Carrier mobility is one the most important parameters of any semiconductor material, determining its suitability for practical use in a variety of electronic devices. In fact, a high charge carrier mobility directly influences the performance of the devices. To examine how fast a photogenerated carrier in arsenic trichalcogenide will be, we used deformation potential theory combined with the effective mass approximation to calculate the electron and hole mobilities of $As_2X_3$ monolayers along $x$ and $y$ directions. Table 1 summarized the calculated elastic modulus ($C_{2D}$), effective mass ($m_i^*$), deformation potential constant ($E_1^j$), and carrier mobility ($\mu_i$) of $As_2X_3$ monolayers. From the table, one can clearly see that for the studied monolayers, electron mobilities are appreciably larger than the hole mobilities along both $x$ and $y$ directions. Such a considerable difference between the electron and hole mibilities may offer desirable conditions for the more effective separation of photogenerated electron and holes, which consequently enhances the efficiency in the hydrogen fuel generation. The electron mobilities along $x$ and $y$ directions were found to be 1075.47 and 2744.14 $cm^2V^{-1}S^{-1}$ for $As_2S_3$, 779.33 and 3384.80 $cm^2V^{-1}S^{-1}$ for $As_2Se_3$, and 433.08 and 368.31 $cm^2V^{-1}S^{-1}$ for $As_2Te_3$, respectively, indicating that arsenic trichalcogenides possess anisotropic electron mobilities as expected from their highly directional porosity. Same behavior is also observed for the hole mobilities along $x$ and $y$ directions. The calculated mobilities are smaller than those of phosphorene ~10000 $cm^2V^{-1}S^{-1}$ [61] while appreciably larger than those of some other 2D materials like $MnPSe_3$ ~626 $cm^2V^{-1}S^{-1}$ [62] and $MoS_2$ ~200 $cm^2V^{-1}S^{-1}$ [63]. According to the results included in table 1, the huge difference between the mobilities of the electrons and holes originates mainly from difference in their calculated deformation potential energies. This can be simply understood by considering the point that for the three monolayers, VBM represents anti-bonding (A-*X*)in-plane σ* states, while CBM is made of bonding (As-*X*)in-plane π and antibonding (As-*X*)vertical σ* states. It is apparent that in-plane σ*(VBM) state is much more sensitive towards the in-plane lattice dilation than in-plane π states(CBM), resulting in much larger deformation potential energies for holes along the two directions.

**Table 1**. Elastic Modulus ($C_{2D}$), effective mass ($m_e^*$, $m_h^*$) of electrons and holes with respect to the free-electron mass ($m_0$), deformation potential constant of VBM and CBM ($E_{VBM}^1$, $E_{CBM}^1$), and mobility ($\mu_e$, $\mu_h$) of electrons and holes along *x* and *y* directions for As$_2$X$_3$ monolayers. All parameters presented in the Table were calculated using HSE06 functional.

|  | Direction | $C_{2D}$(J/m$^2$) | Hole | | | Electron | | |
|---|---|---|---|---|---|---|---|---|
|  |  |  | $m_h^*$ | $E_{VBM}^1$(eV) | $\mu_h$ | $m_e^*$ | $E_{CBM}^1$(eV) | $\mu_e$ |
| As$_2$S$_3$ | *x* | 29.4 | 1.30 | 2.78 | 26.78 | 1.70 | 0.53 | 1075.47 |
|  | *y* | 8.3 | 4.20 | 0.64 | 44.67 | 0.92 | 0.24 | 2744.14 |
| As$_2$Se$_3$ | *x* | 27.5 | 1.56 | 3.64 | 28.89 | 1.91 | 0.56 | 779.33 |
|  | *y* | 9.1 | 0.62 | 2.07 | 75.30 | 0.86 | 0.23 | 3384.80 |
| As$_2$Te$_3$ | *x* | 28.7 | 0.42 | 4.31 | 183.62 | 2.46 | 0.52 | 433.08 |
|  | *y* | 9.9 | 0.44 | 2.60 | 164.81 | 1.82 | 0.39 | 368.31 |

[a]Mobility at 300 K in unit of cm$^2$V$^{-1}$s$^{-1}$.

We next explore the optical properties of single-layer and bilayer As$_2$S$_3$, As$_2$Se$_3$ and As$_2$Te$_3$ according to the BSE+HSE06 method results. Because of the highly anisotropic nature of the considered nanosheets the optical properties were also calculated along the both planar directions. In addition to having an appropriate band-gap value and high carrier mobilities, a semiconductor is expected to also absorb a significant portion of the incident visible light in order to be suitable for photocatalytic activity and water splitting accordingly. We thus examined the photoactivity of arsenic trichalcogenide monolayers and bilayers, by calculating their optical absorption spectra as shown in Fig. 9. It can be seen that As$_2$Se$_3$ and As$_2$Te$_3$ monolayers and bilayers exhibit optical absorption in the visible region. However, although As$_2$S$_3$ monolayer and bilayers show proper band edge positions for the photocatalytic water splitting under different pH conditions, they show very low optical absorption in the visible region, indicating that they might have no photocatalytic activity for solar water splitting. It can be thus concluded that among the all studied nanosheets, As$_2$Se$_3$ nanosheets exhibit the brightest performance for the overall solar water splitting.

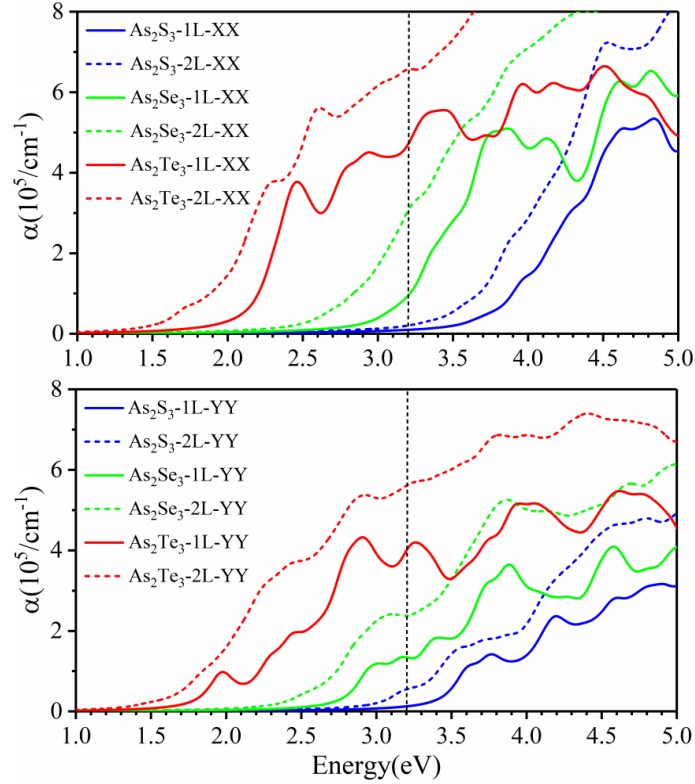

**Fig. 9,** Optical properties of As$_2$X$_3$ nanosheets along the *x* (XX) and *y* (YY) directions predicted by the BSE+HSE06 method.

## 4. Summary

In this work, we explored the mechanical properties, stability, electronic/optical features and photocatalytic characteristics of monoclinic As$_2$X$_3$ (X=S, Se and Te) nanosheets via DFT calculations. While As$_2$S$_3$ and As$_2$Se$_3$ in the bulk form are already available in the nature, we predicted a novel stable As$_2$Te$_3$ 2D lattice, with the similar atomic lattice as that of the monoclinic As$_2$S$_3$ or As$_2$Se$_3$. Desirable thermal and dynamical stability of all studied nanosheets were confirmed by the phonon dispersions and ab-initio molecular dynamics results. In addition, relatively weak interlayer interactions in the multi-layered As$_2$S$_3$ and As$_2$Se$_3$ nanosheets confirm that the exfoliation methods can be employed to fabricate these nanomembranes from their bulk counterparts. Uniaxial tensile simulation results reveal highly anisotropic mechanical properties of As$_2$X$_3$ nanosheets. Notably, our results confirm that stretchability of As$_2$S$_3$ and As$_2$Te$_3$ are by more than two and three folds larger than graphene, respectively. Particularly, As$_2$Te$_3$ nanosheet was found to outperform other known 2D materials with respect to the stretchability. This observation can be used to guide the design and synthesize of other superstretchable 2D materials. HSE06 functional results confirm the semiconducting electronic nature for the monolayer to few-layer and bulk As$_2$X$_3$ lattices. Moreover, our electronic structure results reveal the indirect band-gap in the studied 2D systems, in which decreasing trends in the band-gap values by increasing the thickness exist. Notably, HSE06 estimations confirm that $As_2S_3$ and $As_2Se_3$ nanosheets can exhibit suitable valence and conduction band edge positions for the overall solar water splitting under various environments with different pHs. Therewith, the studied nanosheets were found to yield remarkably high and anisotropic carrier mobilities. It was found that electron mobilities are considerably larger than those for holes. In addition, optical results show that these nanosheets can absorb the visible light. In particular, $As_2Se_3$ nanosheets were found to yield highly promising electronic structure, electron and hole mobilities and optical features for the application in hydrogen fuel production via solar water splitting. Our extensive first-principles results provide a comprehensive vision concerning the stability, mechanical, electronic and optical characteristics of $As_2S_3$, $As_2Se_3$ and $As_2Te_3$ nanosheets. Our results not only suggest these novel 2D semiconductors for the applications in energy storage/conversion, nanooptics and nanoelectronics systems, but may also be used to design novel superstretchable devices.

**Acknowledgment**

B. M. and X. Z. appreciate the funding by the Deutsche Forschungsgemeinschaft (DFG, German Research Foundation) under Germany's Excellence Strategy within the Cluster of Excellence PhoenixD (EXC 2122, Project ID 390833453).

**Appendix A. Supplementary data**

The following are the supplementary data to this article:

**References**

[1] K.S. Novoselov, A.K. Geim, S. V Morozov, D. Jiang, Y. Zhang, S. V Dubonos, I. V Grigorieva, A.A. Firsov, Electric field effect in atomically thin carbon films., Science. 306 (2004) 666–9. doi:10.1126/science.1102896.

[2] A.K. Geim, K.S. Novoselov, The rise of graphene, Nat. Mater. 6 (2007) 183–191. doi:10.1038/nmat1849.

[3] A.H.. Castro Neto, N.M.R.. Peres, K.S.. Novoselov, A.K.. Geim, F. Guinea, The electronic properties of graphene, Rev. Mod. Phys. 81 (2009) 109–162. doi:10.1103/RevModPhys.81.109.

[4] C. Lee, X. Wei, J.W. Kysar, J. Hone, Measurement of the Elastic Properties and Intrinsic Strength of Monolayer Graphene, Science (80-. ). 321 (2008) 385–388.

doi:10.1126/science.1157996.

[5]   A.A. Balandin, S. Ghosh, W. Bao, I. Calizo, D. Teweldebrhan, F. Miao, C.N. Lau, Superior thermal conductivity of single-layer graphene, Nano Lett. 8 (2008) 902–907. doi:10.1021/nl0731872.

[6]   M. Yankowitz, S. Chen, H. Polshyn, Y. Zhang, K. Watanabe, T. Taniguchi, D. Graf, A.F. Young, C.R. Dean, Tuning superconductivity in twisted bilayer graphene, Science (80-. ). (2019). doi:10.1126/science.aav1910.

[7]   M. Liu, X. Yin, E. Ulin-Avila, B. Geng, T. Zentgraf, L. Ju, F. Wang, X. Zhang, A graphene-based broadband optical modulator, Nature. 474 (2011) 64–67. doi:10.1038/nature10067.

[8]   F. Withers, M. Dubois, A.K. Savchenko, Electron properties of fluorinated single-layer graphene transistors, Phys. Rev. B - Condens. Matter Mater. Phys. 82 (2010). doi:10.1103/PhysRevB.82.073403.

[9]   Y. Kubota, K. Watanabe, O. Tsuda, T. Taniguchi, Deep ultraviolet light-emitting hexagonal boron nitride synthesized at atmospheric pressure., Science. 317 (2007) 932–934. doi:10.1126/science.1144216.

[10]  L. Song, L. Ci, H. Lu, P.B. Sorokin, C. Jin, J. Ni, A.G. Kvashnin, D.G. Kvashnin, J. Lou, B.I. Yakobson, P.M. Ajayan, Large scale growth and characterization of atomic hexagonal boron nitride layers, Nano Lett. 10 (2010) 3209–3215. doi:10.1021/nl1022139.

[11]  B. Aufray, A. Kara, Ś. Vizzini, H. Oughaddou, C. Ĺandri, B. Ealet, G. Le Lay, Graphene-like silicon nanoribbons on Ag(110): A possible formation of silicene, Appl. Phys. Lett. 96 (2010). doi:10.1063/1.3419932.

[12]  P. Vogt, P. De Padova, C. Quaresima, J. Avila, E. Frantzeskakis, M.C. Asensio, A. Resta, B. Ealet, G. Le Lay, Silicene: Compelling experimental evidence for graphenelike two-dimensional silicon, Phys. Rev. Lett. 108 (2012). doi:10.1103/PhysRevLett.108.155501.

[13]  E. Bianco, S. Butler, S. Jiang, O.D. Restrepo, W. Windl, J.E. Goldberger, Stability and exfoliation of germanane: A germanium graphane analogue, ACS Nano. 7 (2013) 4414–4421. doi:10.1021/nn4009406.

[14]  D.A. Bandurin, A. V. Tyurnina, G.L. Yu, A. Mishchenko, V. Zólyomi, S. V. Morozov, R.K. Kumar, R. V. Gorbachev, Z.R. Kudrynskyi, S. Pezzini, Z.D. Kovalyuk, U. Zeitler, K.S. Novoselov, A. Patanè, L. Eaves, I. V. Grigorieva, V.I. Fal'ko, A.K. Geim, Y. Cao, High electron mobility, quantum Hall effect and anomalous optical response

in atomically thin InSe, Nat. Nanotechnol. (2016) 1–18. doi:10.1038/nnano.2016.242.

[15] Y. Wang, Y. Ding, Strain-induced self-doping in silicene and germanene from first-principles, Solid State Commun. 155 (2013) 6–11. doi:10.1016/j.ssc.2012.10.044.

[16] B. Radisavljevic, A. Radenovic, J. Brivio, V. Giacometti, A. Kis, Single-layer MoS2 transistors, Nat. Nanotechnol. 6 (2011) 147–50. doi:10.1038/nnano.2010.279.

[17] A.J. Mannix, B. Kiraly, M.C. Hersam, N.P. Guisinger, Synthesis and chemistry of elemental 2D materials, Nat. Rev. Chem. 1 (2017). doi:10.1038/s41570-016-0014.

[18] X.F. Zhou, X. Dong, A.R. Oganov, Q. Zhu, Y. Tian, H.T. Wang, Semimetallic two-dimensional boron allotrope with massless Dirac fermions, Phys. Rev. Lett. 112 (2014). doi:10.1103/PhysRevLett.112.085502.

[19] M.C. Watts, L. Picco, F.S. Russell-Pavier, P.L. Cullen, T.S. Miller, S.P. Bartuś, O.D. Payton, N.T. Skipper, V. Tileli, C.A. Howard, Production of phosphorene nanoribbons, Nature. (2019) 216–220. doi:10.1038/s41586-019-1074-x.

[20] R. Gusmão, Z. Sofer, M. Pumera, Functional Protection of Exfoliated Black Phosphorus by Noncovalent Modification with Anthraquinone, ACS Nano. 12 (2018) 5666–5673. doi:10.1021/acsnano.8b01474.

[21] J. Sturala, Z. Sofer, M. Pumera, Chemistry of Layered Pnictogens: Phosphorus, Arsenic, Antimony, and Bismuth, Angew. Chemie - Int. Ed. (2019) 7551–7557. doi:10.1002/anie.201900811.

[22] D. Jariwala, V.K. Sangwan, L.J. Lauhon, T.J. Marks, M.C. Hersam, Emerging device applications for semiconducting two-dimensional transition metal dichalcogenides, ACS Nano. 8 (2014) 1102–1120. doi:10.1021/nn500064s.

[23] R. Mohammad, D. Kenneth, Q. Shi-Zhang, Advent of 2D Rhenium Disulfide (ReS2): Fundamentals to Applications, Adv. Funct. Mater. 27 (2017) 1606129. doi:10.1002/adfm.201606129.

[24] Y. Shengxue, Y. Yanhan, W. Minghui, H. Chunguang, S. Wanfu, G. Yongji, H. Li, J. Chengbao, Z. Yongzhe, A.P. M., Highly In-Plane Optical and Electrical Anisotropy of 2D Germanium Arsenide, Adv. Funct. Mater. 0 (2018) 1707379. doi:10.1002/adfm.201707379.

[25] E. Liu, Y. Fu, Y. Wang, Y. Feng, H. Liu, X. Wan, W. Zhou, B. Wang, L. Shao, C.-H. Ho, Y.-S. Huang, Z. Cao, L. Wang, A. Li, J. Zeng, F. Song, X. Wang, Y. Shi, H. Yuan, H.Y. Hwang, Y. Cui, F. Miao, D. Xing, Integrated digital inverters based on two-dimensional anisotropic ReS2 field-effect transistors, Nat. Commun. 6 (2015) 6991. doi:10.1038/ncomms7991.

[26] J. Yan, M. Hu, D. Li, Y. He, R. Zhao, X. Jiang, S. Song, L. Wang, C. Fan, A nano- and micro- integrated protein chip based on quantum dot probes and a microfluidic network, Nano Res. 1 (2008) 490–496. doi:10.1007/s12274-008-8052-1.

[27] H. Yang, H. Jussila, A. Autere, H.P. Komsa, G. Ye, X. Chen, T. Hasan, Z. Sun, Optical Waveplates Based on Birefringence of Anisotropic Two-Dimensional Layered Materials, ACS Photonics. 4 (2017) 3023–3030. doi:10.1021/acsphotonics.7b00507.

[28] S. V Badalov, A. Kandemir, H. Sahin, Monolayer AsTe2: Stable Robust Metal in 2D, 1D and 0D, ChemPhysChem. 19 (2018) 2176–2182. doi:10.1002/cphc.201800473.

[29] S. Zhang, S. Guo, Z. Chen, Y. Wang, H. Gao, J. Gómez-Herrero, P. Ares, F. Zamora, Z. Zhu, H. Zeng, Recent progress in 2D group-VA semiconductors: from theory to experiment, Chem. Soc. Rev. 47 (2018) 982–1021. doi:10.1039/C7CS00125H.

[30] J. Sturala, Z. Sofer, M. Pumera, Coordination chemistry of 2D and layered gray arsenic: photochemical functionalization with chromium hexacarbonyl, NPG Asia Mater. 11 (2019) 42. doi:10.1038/s41427-019-0142-x.

[31] M. Šiškins, M. Lee, F. Alijani, M.R. van Blankenstein, D. Davidovikj, H.S.J. van der Zant, P.G. Steeneken, Highly Anisotropic Mechanical and Optical Properties of 2D Layered As2S3 Membranes, ACS Nano. (2019). doi:10.1021/acsnano.9b06161.

[32] B. Mortazavi, M. Shahrokhi, G. Cuniberti, X. Zhuang, Two-Dimensional SiP, SiAs, GeP and GeAs as Promising Candidates for Photocatalytic Applications, Coatings. 9 (2019) 522. doi:10.3390/coatings9080522.

[33] G. Kresse, J. Furthmüller, Efficiency of ab-initio total energy calculations for metals and semiconductors using a plane-wave basis set, Comput. Mater. Sci. 6 (1996) 15–50. doi:10.1016/0927-0256(96)00008-0.

[34] G. Kresse, J. Furthmüller, Efficient iterative schemes for ab initio total-energy calculations using a plane-wave basis set, Phys. Rev. B. 54 (1996) 11169–11186. doi:10.1103/PhysRevB.54.11169.

[35] G. Kresse, From ultrasoft pseudopotentials to the projector augmented-wave method, Phys. Rev. B. 59 (1999) 1758–1775. doi:10.1103/PhysRevB.59.1758.

[36] J. Perdew, K. Burke, M. Ernzerhof, Generalized Gradient Approximation Made Simple., Phys. Rev. Lett. 77 (1996) 3865–3868. doi:10.1103/PhysRevLett.77.3865.

[37] S. Grimme, Semiempirical GGA-type density functional constructed with a long-range dispersion correction, J. Comput. Chem. 27 (2006) 1787–1799. doi:10.1002/jcc.20495.

[38] H. Monkhorst, J. Pack, Special points for Brillouin zone integrations, Phys. Rev. B. 13 (1976) 5188–5192. doi:10.1103/PhysRevB.13.5188.

[39] A.V.K. and O.A.V. and A.F.I. and G.E. Scuseria, Influence of the exchange screening parameter on the performance of screened hybrid functionals, J. Chem. Phys. 125 (2006) 224106. doi:10.1063/1.2404663.

[40] A. Togo, I. Tanaka, First principles phonon calculations in materials science, Scr. Mater. 108 (2015) 1–5. doi:10.1016/j.scriptamat.2015.07.021.

[41] S. Albrecht, L. Reining, R. Del Sole, G. Onida, Ab Initio Calculation of Excitonic Effects in the Optical Spectra of Semiconductors, Phys. Rev. Lett. 80 (1998) 4510–4513. doi:10.1103/PhysRevLett.80.4510.

[42] J. Bardeen, W. Shockley, Deformation potentials and mobilities in non-polar crystals, Phys. Rev. (1950). doi:10.1103/PhysRev.80.72.

[43] N. MORIMOTO, THE CRYSTAL STRUCTURE OF ORPIMENT ($As_2S_3$) REFINED, Mineral. J. 1 (1954) 160–169. doi:10.2465/minerj1953.1.160.

[44] B. Silvi, A. Savin, Classification of Chemical-Bonds Based on Topological Analysis of Electron Localization Functions, Nature. 371 (1994) 683–686. doi:10.1038/371683a0.

[45] Z. Zhu, J. Guan, D. Liu, D. Tománek, Designing Isoelectronic Counterparts to Layered Group V Semiconductors, ACS Nano. (2015). doi:10.1021/acsnano.5b02742.

[46] W. Wang, S. Dai, X. Li, J. Yang, D.J. Srolovitz, Q. Zheng, Measurement of the cleavage energy of graphite, Nat. Commun. (2015). doi:10.1038/ncomms8853.

[47] S. Zhao, Z. Li, J. Yang, Obtaining two-dimensional electron gas in free space without resorting to electron doping: An electride based design, J. Am. Chem. Soc. (2014). doi:10.1021/ja5065125.

[48] F. Liu, P. Ming, J. Li, Ab initio calculation of ideal strength and phonon instability of graphene under tension, Phys. Rev. B - Condens. Matter Mater. Phys. 76 (2007). doi:10.1103/PhysRevB.76.064120.

[49] Q. Peng, W. Ji, S. De, Mechanical properties of the hexagonal boron nitride monolayer: Ab initio study, Comput. Mater. Sci. 56 (2012) 11–17. doi:10.1016/j.commatsci.2011.12.029.

[50] B. Mortazavi, M. Shahrokhi, X. Zhuang, T. Rabczuk, Boron-graphdiyne: A superstretchable semiconductor with low thermal conductivity and ultrahigh capacity for Li, Na and Ca ion storage, J. Mater. Chem. A. 6 (2018) 11022–11036. doi:10.1039/c8ta02627k.

[51] L.-B. Shi, S. Cao, M. Yang, Strain behavior and Carrier mobility for novel two-dimensional semiconductor of GeP: First principles calculations, Phys. E Low-

Dimensional Syst. Nanostructures. 107 (2019) 124–130. doi:10.1016/J.PHYSE.2018.11.024.

[52] A. Bafekry, C. Stampfl, saber Shayesteh, A first-principles study of C3N nanostructures: Control and engineering of the electronic and magnetic properties of nanosheets, tubes and ribbons, ChemPhysChem. n/a (2019). doi:10.1002/cphc.201900852.

[53] A. Bafekry, C. Stampfl, M. Ghergherehchi, S. Farjami Shayesteh, A first-principles study of the effects of atom impurities, defects, strain, electric field and layer thickness on the electronic and magnetic properties of the C2N nanosheet, Carbon N. Y. 157 (2020) 371–384. doi:https://doi.org/10.1016/j.carbon.2019.10.038.

[54] A. Bafekry, S. Farjami Shayesteh, M. Ghergherehchi, F.M. Peeters, Tuning the bandgap and introducing magnetism into monolayer BC3 by strain/defect engineering and adatom/molecule adsorption, J. Appl. Phys. 126 (2019) 144304. doi:10.1063/1.5097264.

[55] A. Bafekry, S.F. Shayesteh, F.M. Peeters, Two-dimensional carbon nitride (2DCN) nanosheets: Tuning of novel electronic and magnetic properties by hydrogenation, atom substitution and defect engineering, J. Appl. Phys. 126 (2019) 215104. doi:10.1063/1.5120525.

[56] A. Bafekry, M. Ghergherehchi, S. Farjami Shayesteh, F.M. Peeters, Adsorption of molecules on C3N nanosheet: A first-principles calculations, Chem. Phys. 526 (2019) 110442. doi:https://doi.org/10.1016/j.chemphys.2019.110442.

[57] L.-B. Shi, S. Cao, M. Yang, Q. You, K.-C. Zhang, Y. Bao, Y.-J. Zhang, Y.-Y. Niu, P. Qian, Theoretical prediction of intrinsic electron mobility of monolayer InSe: first-principles calculation, J. Phys. Condens. Matter. 32 (2019) 65306. doi:10.1088/1361-648x/ab534f.

[58] L.-B. Shi, M. Yang, S. Cao, Q. You, Y.-Y. Niu, Y.-Z. Wang, Elastic behavior and intrinsic carrier mobility for monolayer SnS and SnSe: First-principles calculations, Appl. Surf. Sci. 492 (2019) 435–448. doi:https://doi.org/10.1016/j.apsusc.2019.06.211.

[59] Y. Tachibana, L. Vayssieres, J.R. Durrant, Artificial photosynthesis for solar water-splitting, Nat. Photonics. (2012) 511–518. doi:10.1038/nphoton.2012.175.

[60] N.S. Lewis, D.G. Nocera, Powering the planet: Chemical challenges in solar energy utilization, Proc. Natl. Acad. Sci. 103 (2006) 15729– 15735. doi:10.1073/pnas.0603395103.

[61] J. Qiao, X. Kong, Z.-X. Hu, F. Yang, W. Ji, High-mobility transport anisotropy and

linear dichroism in few-layer black phosphorus, Nat. Commun. 5 (2014) 4475. doi:10.1038/ncomms5475.

[62] X. Zhang, X. Zhao, D. Wu, Y. Jing, Z. Zhou, MnPSe3 Monolayer: A Promising 2D Visible-Light Photohydrolytic Catalyst with High Carrier Mobility, Adv. Sci. 3 (2016) 1600062. doi:10.1002/advs.201600062.

[63] Y. Cai, G. Zhang, Y.-W. Zhang, Polarity-Reversed Robust Carrier Mobility in Monolayer MoS2 Nanoribbons, J. Am. Chem. Soc. 136 (2014) 6269–6275. doi:10.1021/ja4109787.

# Supporting Information

# As$_2$S$_3$, As$_2$Se$_3$ and As$_2$Te$_3$ nanosheets: Superstretchable semiconductors with anisotropic carrier mobilities and optical properties

Bohayra Mortazavi[*a], Fazel Shojaei[b], Maryam Azizi[b], Timon Rabczuk[c] and Xiaoying Zhuang[c]

[a]Institute of Continuum Mechanics, Leibniz Universität Hannover, Appelstraße 11, 30157 Hannover, Germany.
[b]Schoolof Nano Science, Institute for Research in Fundamental Sciences (IPM), 19395-5531 Tehran, Iran.
[c]College of Civil Engineering, Department of Geotechnical Engineering, Tongji University, Shanghai, China.

*E-mail:  bohayra.mortazavi@gmail.com

1. Atomic structures in VASP POSCAR format.

2- AIMD results for the thermal stability.

3- PBE results for the electronic band structures of monolayers with and without SOC.

4- HSE06 results for the electronic band structures of bulk systems.

5- Charge density distributions of VBM and CBM states of As$_2$Se$_3$ and As$_2$Te$_3$ monolayers.

1. Atomic structures in VASP POSCAR format.

**As2S3 Single-layer**
1.00000000000000
     4.4565478770785676    0.0000000000000000    0.0000000000000000
     0.0000000000000000   11.3586428131330397    0.0000000000000000
     0.0000000000000000    0.0000002190398780   16.0000000000000000
   As    S
    4     6
Direct
  0.1470184054197524  0.9776839827368167  0.6887287834054856
  0.6470185841413943  0.4776834832616674  0.7790583578762966
  0.1469763350112658  0.2662003801634338  0.6887857330475007
  0.6469778898592778  0.7662009364405205  0.7790015304777591
  0.9866464694417696  0.9089773362034305  0.8151821001991848
  0.4866404976576784  0.4089780183359392  0.6526051852682671
  0.4866654382906487  0.8349020606950012  0.6525363747958466
  0.9866695313801641  0.3349025991376285  0.8152510745476576
  0.4455842341161372  0.1219276070633198  0.7472568477822522
  0.9455816968064539  0.6219274337906382  0.7205293878497953

**As2S3 Double-layer**
   1.00000000000000
     4.3783708174226312    0.0000000000000000    0.0000000000000000
     0.0000000000000000   11.4586199205554848    0.0000000000000000
     0.0000000000000000    0.0000000000000000   32.0000000000000000
   As    S
    8    12
Direct
  0.1476807229525636  0.9764001531148351  0.0821023790869207
  0.8866680497627411  0.0238772665910248  0.2827966721130776
  0.3845476789411656  0.5206238787009527  0.2366695950008403
  0.6498024329920845  0.4796540133580149  0.1282295078658254
  0.1467759572848243  0.2626517593597238  0.0855045099745444
  0.8875738457144559  0.7376260487057881  0.2793945690254456
  0.3915113425880539  0.2377367875308832  0.2388101260614465
  0.6428382800197165  0.7625410205346216  0.1260890085718868
  0.0440406091334183  0.0980180727110716  0.2204024687257566
  0.9903087721601976  0.9022595044164765  0.1444966103075791
  0.4903251986279764  0.4035568312536605  0.0659180015128657
  0.5440247331297903  0.5967210188085651  0.2989811172537996
  0.4900123471912655  0.8351999193980839  0.0626270828955719
  0.5443371209495235  0.1650779411413145  0.3022720080377560
  0.0433574409834475  0.6631311655411537  0.2163932380246875
  0.9909910420252519  0.3371466214819690  0.1485058568753162
  0.5871515494438349  0.8820153259917826  0.2504388159414133
  0.4471965794790087  0.1182625660671708  0.1144602551252533
  0.9482540911220213  0.6192453168391492  0.0965388319792146
  0.0860949520725384  0.3810324912263557  0.2683602390874521

**1.2 AS2Se3 Single-layer**
```
   1.00000000000000
     4.4329380980850432    0.0000000000000000    0.0000000000000000
     0.0000000000000000   12.0962864784322601    0.0000000000000000
     0.0000000000000000    0.0000000000000000   16.0000000000000000
   As   Se
    4    6
Direct
  0.1391577113197329  0.9820766577777462  0.7094611954566901
  0.6388662362776714  0.4821039964613121  0.7943627316320843
  0.1380285587681058  0.2619229220143922  0.7104187158773172
  0.6384230657526899  0.7619271393894564  0.7934272775622162
  0.9870913661965659  0.9053635920966059  0.8422273659281956
  0.4867017030485243  0.4053654239248701  0.6616115766589203
  0.4874966120760078  0.8385578428978278  0.6605154279984546
  0.9870802902607068  0.3385552440965114  0.8433304986732750
  0.4632864370828500  0.1217573585943228  0.7768619366632663
  0.9635019706516559  0.6217207437508208  0.7269983208620872
```

**As2Se3 Double-layer**
```
   1.00000000000000
     4.3601426900604530    0.0000000000000000    0.0000000000000000
     0.0000000000000000   12.1836960300425350    0.0000000000000000
     0.0000000000000000    0.0000000000000000   30.0000000000000000
   As   Se
    8   12
Direct
  0.1290778006839928  0.9833715911468249  0.0842022324832422
  0.9081803132203292  0.0167657163496198  0.2990605508024734
  0.4066191997568521  0.5126733100829902  0.2499000769787249
  0.6306402455322430  0.4874645065007194  0.1333627531641222
  0.1238445915157327  0.2633637581923829  0.0903481208382409
  0.9134145824649467  0.7367740192181322  0.2929146667331913
  0.4201497822435694  0.2376830360935972  0.2548118460142340
  0.6171091202118224  0.7624547608624493  0.1284510096643395
  0.0565120395877425  0.0985921407657860  0.2289730677670713
  0.9807468077818502  0.9015453429689680  0.1542897282686516
  0.4780413743397016  0.4048923182730638  0.0634930817049612
  0.5592178900771164  0.5952454199642568  0.3197697654378980
  0.4818282966266096  0.8422254200918825  0.0569453659559240
  0.5554305023802470  0.1579124161194831  0.3263174471512134
  0.0575561271098335  0.6560601433763223  0.2217837937451063
  0.9797016087844457  0.3440776340341857  0.1614790278620407
  0.5826787251213347  0.8789199206051984  0.2595453877073876
  0.4545787294883667  0.1212179351517057  0.1237173998640377
  0.9517384079984404  0.6219949765892182  0.0938867201642558
  0.0855198546809182  0.3781428008212898  0.2893760333357500
```

**1.3 As2Te3 Single-layer**
```
   1.00000000000000
     4.4586902569312423    0.0000000000000000    0.0000000000000000
     0.0000000000000000   13.1148798657304209    0.0000000000000000
     0.0000000000000000    0.0000000000000000   16.0000000000000000
   As   Te
    4    6
Direct
  0.1236157189792877  0.9858096511681538  0.6945001095466310
  0.6236253649356485  0.4858073993371050  0.7732886945177313
  0.1236754961633295  0.2580628461431047  0.6944478802677878
  0.6236556575417254  0.7580628167288381  0.7733367520809367
  0.9927269230368125  0.8999795028050741  0.8364194578146922
  0.4927573983522038  0.3999735791446213  0.6313692101721884
  0.4927130572048100  0.8439005761353755  0.6314245877740845
  0.9927175882574986  0.3438908626917723  0.8363623665739865
  0.4801517226618088  0.1219475665441792  0.7726713461316166
  0.9801402532005312  0.6219488813383904  0.6951149106828183
```

**As2Te3 Double-layer**
```
   1.00000000000000
     4.3911023379122733    0.0000000000000000    0.0000000000000000
     0.0000000000000000   13.1851312463480443    0.0000000000000000
     0.0000000000000000    0.0000000000000000   32.0000000000000000
   As   Te
    8   12
Direct
  0.1151653011780240  0.9892635289554192  0.0718186530081866
  0.9339983631709075  0.0106535007461943  0.2872536110251431
  0.4336164326067333  0.5051965916502326  0.2360555481191540
  0.6155485924441280  0.4947210170604635  0.1230167795141739
  0.0967919078861499  0.2660654309349866  0.0836020412872074
  0.9523728946241050  0.7338520519631615  0.2754702624794589
  0.4585572468914805  0.2358143343944263  0.2445340953978032
  0.5906073069614735  0.7641031485037288  0.1145382560688617
  0.0646752919648605  0.0981970756723882  0.2157250009287850
  0.9844890854357757  0.9017201197188456  0.1433472869712134
  0.4782063243940797  0.4059569825600278  0.0525161597865846
  0.5709584910647933  0.5939605183544145  0.3065561757800808
  0.4913627369292897  0.8484837489281887  0.0415547885851879
  0.5578016075511414  0.1514338597825073  0.3175175270814776
  0.0654214515354425  0.6477214999464991  0.2029214060864675
  0.9837419382589587  0.3521959650107362  0.1561508976801989
  0.5799519671367364  0.8762874024763363  0.2430955332601172
  0.4692110693141061  0.1236301972639069  0.1159767625398800
  0.9648484226824081  0.6245952937041608  0.0790756185371393
  0.0843156079587004  0.3753222430921141  0.2799966296295295
```

2- AIMD results for the thermal stability.

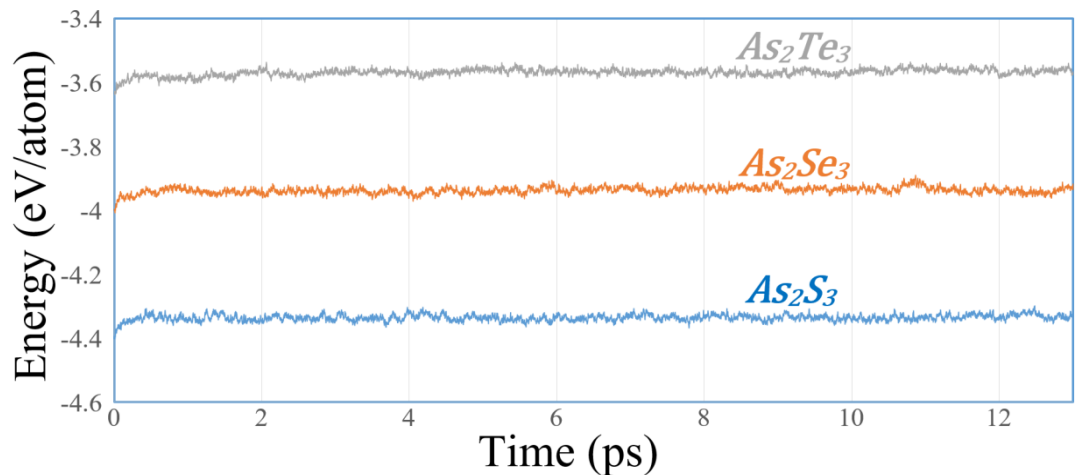

**Fig. S1**, Fluctuation of per atoms energy during the AIMD simulations at 500 K.

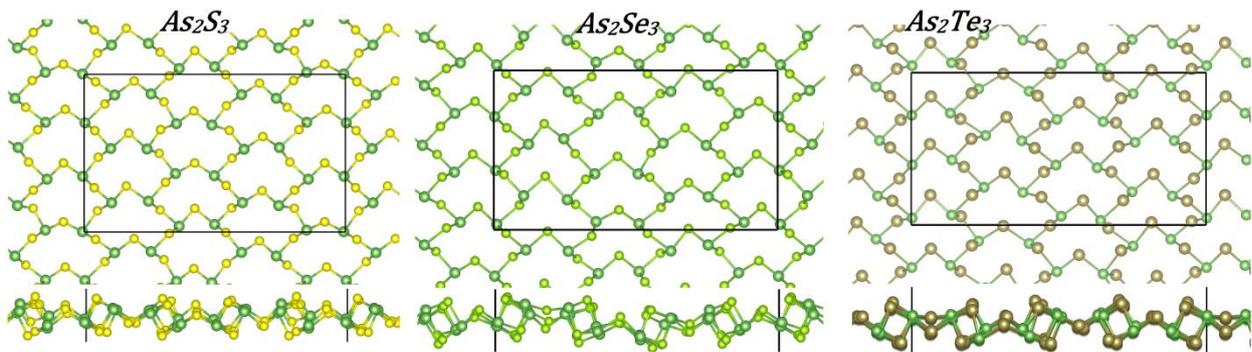

**Fig. S2**, Top and side views of studied monolayers after the AIMD simulations for 13 ps.

**3-** PBE results for the electronic band structures of monolayers with and without SOC.

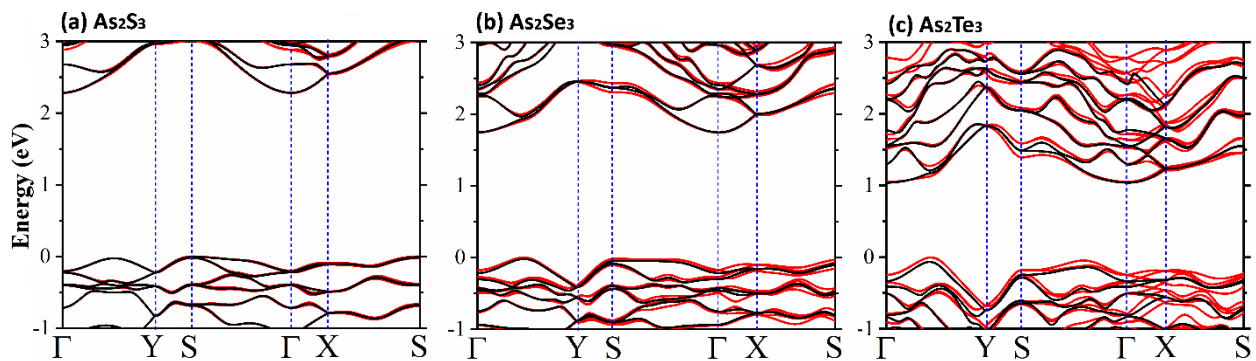

**Fig. S3**, PBE results for the electronic band structures of $As_2S_3$, $As_2Se_3$ and $As_2Te_3$ monolayer with SOC (red lines) and without SOC (black lines).

**4-** HSE06 results for the electronic band structures of bulk systems.

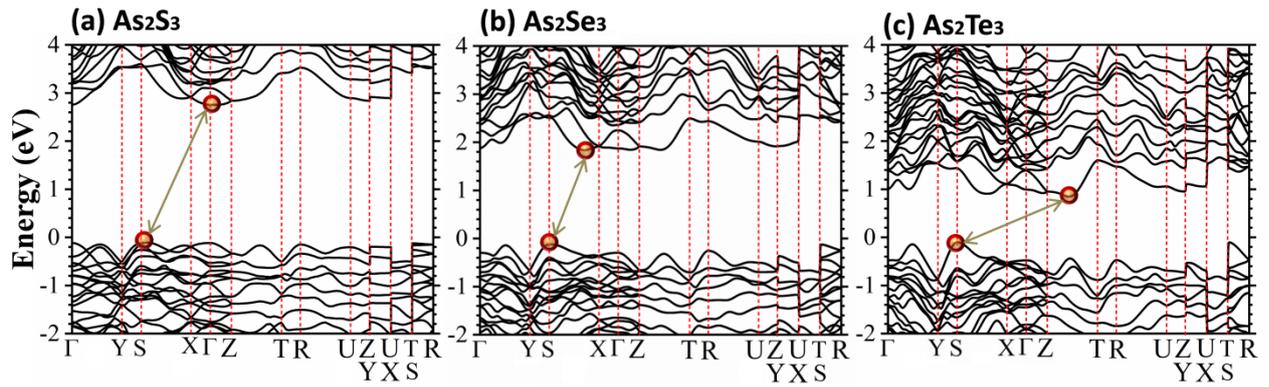

**Fig. S4**, HSE06 results for the electronic band structures of bulk $As_2S_3$, $As_2Se_3$ and $As_2Te_3$.

**5-** Charge density distributions of VBM and CBM states of $As_2Se_3$ and $As_2Te_3$ monolayers.

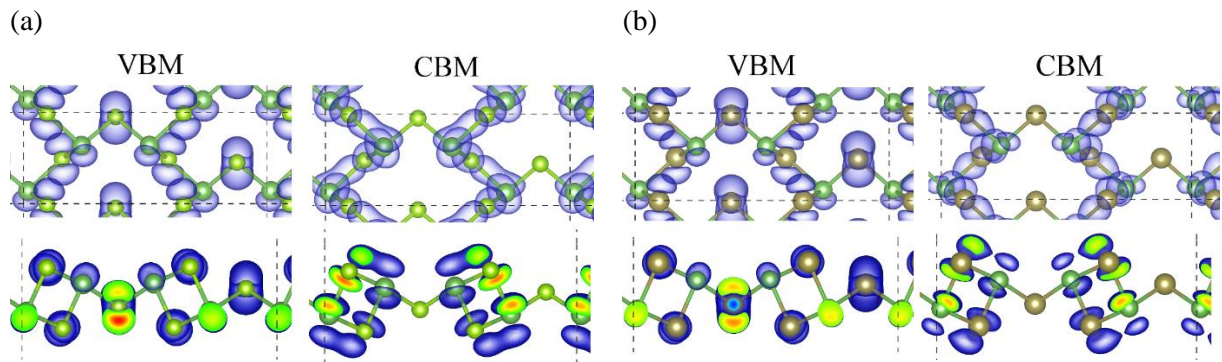

**Fig. S5**, HSE06 calculated charge density distributions of VBM and CBM states of (a) $As_2Se_3$ and (b) $As_2Te_3$ monolayers. The iso-surface value is set to 0.003 e/Å$^3$.